\documentclass[prd,showpacs,preprintnumbers,amsmath,amssymb,11pt]{revtex4}
\usepackage{graphics}
\usepackage[utf8]{inputenc}
\usepackage{epsfig}
\usepackage{subfigure}
\usepackage{dcolumn}% Align table columns on decimal point
\usepackage{bm}% bold math
\usepackage{color}

\begin{document}

\title{Exploring the $\Upsilon(4S,5S,6S) \to h_b(1P)\eta$ hidden-bottom hadronic transitions  }

\author{Yawei Zhang, and Gang Li\footnote{gli@mail.qfnu.edu.cn}
} 

\affiliation{School of Physics and Physical Engineering, Qufu Normal University, Qufu 273165, China}

\begin{abstract}
Recently,   Belle Collaboration has   reported the measurement   of the spin-flipping   transition $\Upsilon(4S) \to h_b(1P)\eta$ with an unexpectedly large branching ratio:  $\mathcal{B}(\Upsilon(4S) \to h_b(1P)\eta) =(2.18\pm 0.11\pm 0.18)\times 10^{-3}$. Such a large branching fraction contradicts with  the  anticipated   suppression for  the spin flip. In this work, we examine the  effects induced by intermediate bottomed meson loops and point out that these effects are significantly important.  Using the effective Lagrangian approach (ELA), we find the experimental data on $\Upsilon(4S) \to h_b(1P)\eta$ can be accommodated with the reasonable  inputs. We then  explore  the decays $\Upsilon(5S,6S)\to h_b(1P)\eta$  and find that these two channels also have sizable branching fractions.  We also calculate these these processes in the framework of nonrelativistic effective
field theory (NREFT).  For the decays $\Upsilon(4S) \to h_b(1P) \eta$, the NREFT results are at the same order of magnitude but smaller than the ELA results by  a factor of $2$ to $5$. For the decays $\Upsilon(5S, 6S) \to h_b(1P) \eta$  the NREFT results are smaller than the ELA results by approximately one order of magnitude. We suggest future experiment Belle-II to search for the $\Upsilon(5S, 6S)\to h_b(1P) \eta$ decays  which will be helpful to  understand   the transition  mechanism.
\end{abstract}

\date{\today}


%14.40.Rt Exotic mesons

%13.75.Lb Meson-meson interactions

%13.20.Gd Decays of J/\psi, and other quarkonia

%14.40.Pq Heavy quarkonia

%14.40.Lb Charmed mesons

\maketitle

\section{Introduction}
\label{sec:introduction}

In recent years   bottomonium  transitions with  an $\eta$ meson or two pions  in the final state have been extensively  studied on the experimental side~\cite{Aubert:2008az,Collaboration:2011gja,Belle:2011aa,Abe:2007tk,Sokolov:2009zy,BABAR:2011ab,Tamponi:2015xzb}. In 2008, the
BaBar collaboration first observed an enhancement for  the transition $\Upsilon(4S) \to \Upsilon(1S)\eta$  compared to the   dipion transition~\cite{Aubert:2008az}. In 2011, two charged bottomoniumlike structures $Z_b^{\pm}(10610)$ and $Z_b^{\pm}(10650)$ were observed by the Belle Collaboration in the $\pi^{\pm}\Upsilon(nS)$ and $\pi^{\pm}h_b$ invariant mass spectra of $\Upsilon(5S)\to \Upsilon(nS)\pi^+\pi^-$ and $h_b(mP)\pi^+\pi^-$ decays\cite{Collaboration:2011gja,Belle:2011aa}. In 2015, the Belle Collaboration has measured for the first time the branching fraction $\mathcal{B}(\Upsilon(4S) \to h_b(1P)\eta) =(2.18\pm 0.11\pm 0.18)\times 10^{-3}$~\cite{Tamponi:2015xzb}.  This value is   anomalously large since  one would expect a  power suppression for  the transitions with the spin flip~\cite{Kuang:2006me,Voloshin:2007dx}.

A low-lying heavy quarkonium system is expected to be compact and nonrelativistic, so the QCD multipole expansion (QCDME)~\cite{Kuang:2006me,Voloshin:2007dx,Eichten:2007qx} can be applied to explore   the  hadronic transitions. For the excited states that  lie above   open flavor thresholds, QCDME  might be  problematic due to the coupled channel effects.  Several possible  new mechanisms have been proposed in order to explain the anomalous decay widths of $\Upsilon(4S) \to h_b(1P)\eta$.
For instance a nonrelativistic effective
field theory (NREFT) is used in Ref.~\cite{Guo:2010ca}, where  the branching ratio can  reach  the order of $10^{-3}$. It has been noticed for a long time that the intermediate meson loop (IML)  is one prominent nonperturbative mechanism in   hadronic transitions~\cite{Lipkin:1986bi,Lipkin:1988tg,Moxhay:1988ri}.
In recent years, this mechanism has been successfully  applied  to study the production and decays of ordinary and  exotic states~\cite{Liu:2013vfa,Guo:2013zbw,Wang:2013hga,Cleven:2013sq,Chen:2011pv,Li:2012as,Li:2013yla,Voloshin:2013ez,Voloshin:2011qa,Bondar:2011ev,Chen:2011pu,Chen:2012yr,Chen:2013bha,Li:2015uwa,Li:2014gxa,Li:2014uia,Li:2013jma,Li:2013zcr,Li:2011ssa,Guo:2010ak,Wu:2016ypc,Wu:2016dws,Liu:2016xly,Li:2014pfa,Yuan-Jiang:2010cna,Zhao:2013jza,Li:2013xia,Wang:2012mf,Zhang:2009kr,Li:2007xr,Li:2007ky},  and a global agreement with experimental data is  found. This  approach has also been extensively used to study the $\Upsilon(4S, 5S,6S)$ hidden bottomonium decays~\cite{Meng:2007tk,Meng:2008dd,Meng:2008bq,Ke:2010aw,Chen:2011jp,Chen:2014ccr,Wang:2016qmz}.
In this work, we will investigate the process $\Upsilon(4S,5S,6S) \to h_b(1P) \eta$ via IML model.  As we will show in the following the experimental data on $\Upsilon(4S) \to h_b(1P)\eta$ can be accommodated in this approach. We then predict the branching ratios of the decays $\Upsilon(5S,6S)\to h_b(1P)\eta$  and find that they are measurable in future.

The rest of this paper is organized as follows.
We will first introduce the effective Lagrangian for our calculation in Sec.~\ref{sec:formula} and calculate the IML contributions to decay widths. Then, we will present our numerical results in Sec.~\ref{sec:results}. A brief summary will be given in Sec.~\ref{sec:summary}.

%%%%%%%%%%%%%%%%%%%%%
\section{Radiative decays}
\label{sec:formula}
%%%%%%%%%%%%%%%%%%%%%

%%%%%%%%%%%%%%%%%%%%%%%%%%%%%%
%%%%%%%%%%%%%%%%%%%%%%%%%%%%%%
\begin{figure}[hbt]
\begin{center}
\includegraphics[width=1.0\textwidth]{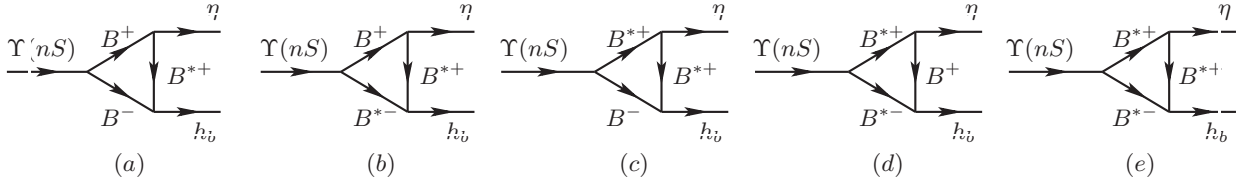}
\vglue-0mm\caption{The hadron-level diagrams for $\Upsilon(4S, 5S, 6S) \to  h_b(1P) \eta $
via charged intermediate bottomed meson loops. Similar diagrams for neutral and strange intermediate bottomed meson loops. } \label{fig:loops}
\end{center}
\end{figure}
%%%%%%%%%%%%%%%%%%%%%%%%%%%%%%
%%%%%%%%%%%%%%%%%%%%%%%%%%%%%%
Generally speaking, all the possible intermediate meson loops should be included in the calculation. In reality, we only pick up the leading order contributions as a reasonable approximation due to the breakdown of the local quark-hadron duality~\cite{Lipkin:1986av,Lipkin:1986bi}.
In this work, we consider the IML illustrated in Fig.~\ref{fig:loops} as the leading order contributions of $\Upsilon(4S,5S,6S) \to h_b(1P) \eta$. To calculate these  diagrams, we need the effective Lagrangians to derive  the couplings. Based on the  heavy quark symmetry and chiral symmetry~\cite{Colangelo:2003sa,Casalbuoni:1996pg}, the Lagrangian
for the S- and P-wave bottomonia at leading order is given as
\begin{eqnarray} \label{eq:S-P-bottomnium}
\mathcal{L}_1 &=& ig_1Tr[P_{b{\bar b}}^\mu {\bar H}_{2i}\gamma _\mu {\bar H }_{1i}]++\mathrm{H.c.} \, , \\
\mathcal{L}_2 &=& g_2 Tr [R_{b{\bar b}}{\bar H}_{2i}\overleftrightarrow{\partial}_\mu\gamma^\mu {\bar H}_1] +\mathrm{H.c.} \, .
\end{eqnarray}
The S-wave bottomonium doublet and P-wave bottomonium multiplet states are
expressed as
\begin{eqnarray}
R_{b{\bar b}} &=& \frac{1 + \not v}{2} \left( \Upsilon^\mu\gamma_\mu-\eta_b\gamma_5 \right)
\frac{1-\not v}{2} \ , \\
P_{b{\bar b}}^\mu &=& \frac{1 + \not v}{2} \left( \chi_{b2}^{\mu
\alpha }\gamma_\alpha +\frac{1}{{\sqrt 2}}\varepsilon^{\mu \nu \alpha
\beta }v_\alpha \gamma_\beta \chi _{b1\nu}+\frac{1}{{\sqrt 3}}(
\gamma^\mu -v^\mu ) \chi_{b0}+h_b^\mu \gamma_5 \right)
\frac{1-\not v}{2} \ ,
\end{eqnarray}
where $\Upsilon$ and $\eta_b$ are the S-wave bottomonium fields. The $h_b$ and $\chi_{bJ}$ (J=0,1,2) are the P-wave bottomonium fields. The  $v^\mu$ is the $4$-velocity of these bottomonium states.

The bottomed and anti-bottomed meson triplet read
\begin{eqnarray}
H_{1i} &=& \frac{1+\not v}{2} \left[ {\cal B}_{i}^{\ast \mu }\gamma _{\mu
}-{\cal B}_{i}\gamma _{5}\right] \, ,  \\
H_{2i} &=&\left[ {\bar{ {\cal B}}}_{i}^{\ast \mu }\gamma _{\mu }- {\bar {\cal B} }_{i}\gamma
_{5}\right] \frac{1-\not v}{2} \ , \\
{\bar H}_{1i,2i}&=&\gamma^0  H_{1i,2i}^\dag \gamma^0 ,
\end{eqnarray}
where ${\cal B}$ and ${\cal B}^{\ast }$ denote the pseudoscalar and vector bottomed meson fields, respectively, i.e. ${\cal B}^{( \ast) }=\left( B^{+(\ast
) },B^{0( \ast ) },B_{s}^{0( \ast ) }\right)$. $v^\mu$ is the $4$-velocity of the bottomed mesons. $\varepsilon_{\mu \nu
\alpha \beta }$ is the antisymmetric Levi-Civita tensor and $\varepsilon_{0123}=+1$.

Consequently, the relevant effective Lagrangian for S-wave $\Upsilon(nS)$ and P-wave $h_b(1P)$ read
\begin{eqnarray}
\mathcal{L}_{\Upsilon(nS) {\cal B}^{(*)} {\cal B}^{(*)}} &=&
ig_{\Upsilon {\cal B}{\cal B}} \Upsilon_{\mu} (\partial^\mu {\cal B} \bar{{\cal B}}- {\cal B}
\partial^\mu \bar{{\cal B}})-g_{\Upsilon {\cal B}^* {\cal B}} \varepsilon_{\mu \nu
\alpha \beta}
\partial^{\mu} \Upsilon^{\nu} (\partial^{\alpha} {\cal B}^{*\beta} \bar{{\cal B}}
 + {\cal B} \partial^{\alpha}
\bar{{\cal B}}^{*\beta})\nonumber\\
&&-ig_{\Upsilon {\cal B}^* {\cal B}^*} \big\{
\Upsilon^\mu (\partial_{\mu} {\cal B}^{* \nu} \bar{{\cal B}}^*_{\nu}
-{\cal B}^{* \nu} \partial_{\mu}
\bar{{\cal B}}^*_{\nu})+ (\partial_{\mu} \Upsilon_{\nu} {\cal B}^{* \nu} -\Upsilon_{\nu}
\partial_{\mu} {\cal B}^{* \nu}) \bar{{\cal B}}^{* \mu}  \nonumber\\
&& +
{\cal B}^{* \mu}(\Upsilon^\nu \partial_{\mu} \bar{{\cal B}}^*_{\nu} -
\partial_{\mu} \Upsilon^\nu \bar{{\cal B}}^*_{\nu})\big\}, \label{eq:h1} \\
\mathcal{L}_{h_b {\cal B}^{( \ast ) }{\cal B}^{( \ast)
}} &=& g_{h_b {\cal B}^\ast {\cal B} }h_b^\mu \left( {\cal B} {\bar{\cal B}}_\mu^\ast + {\cal B}_\mu^\ast {\bar{ \cal B}} \right) +ig_{h_b {\cal B}^\ast {\cal B}^\ast }\varepsilon_{\mu \nu
\alpha \beta }\partial^{\mu }h_b^\nu {\cal B}^{\ast\alpha}{\bar{\cal B}}^{\ast\beta } \, , \label{eq:h2}
\end{eqnarray}
where the coupling constants will be determined later.

The effective Lagrangian for a  light pseudoscalar meson coupled to bottomed mesons pair can be constructed using  the heavy quark symmetry and chiral
symmetry~\cite{Casalbuoni:1996pg,Colangelo:2003sa,Cheng:2004ru}
\begin{equation}
{\mathcal L}_{{\cal B}^{(\ast )}{\cal B}^{(\ast)} {\mathcal P}}=-ig_{{\cal B}^{\ast }{\cal B}
{\mathcal P}}\left( {\cal B}^i \partial^\mu {\mathcal P}_{ij} {\cal B}_\mu^{\ast
j\dagger }-{\cal B}_\mu^{\ast i}\partial^\mu {\mathcal P}_{ij} {\cal B}^{j \dag}\right) +\frac{1}{2}g_{{\cal B}^\ast B^\ast {\mathcal P}}\varepsilon _{\mu
\nu \alpha \beta }{\cal B}_i^{\ast \mu }\partial^\nu {\mathcal P}_{ij}  {\overset{
\leftrightarrow }{\partial }}{\!^{\alpha }} {\cal B}_j^{\ast \beta\dag },
\end{equation}
where ${\mathcal P}$ is  a $3\times 3$ matrix for the pseudoscalar octet. The physical states $\eta$ is the linear combinations of $n{\bar n} = ({u\bar u} + {d\bar d})/\sqrt{2}$ and $s\bar{s}$ with the mixing scheme:
\begin{eqnarray}
|\eta\rangle = \cos\alpha_P|n\bar n\rangle -\sin\alpha_P |s\bar s \rangle.
\end{eqnarray}
The mixing  angle  is given as  $\alpha_P \simeq \theta_P + \arctan \sqrt{2}$, where  the empirical value for the  $\theta_P$ should be in the range $-24.6^\circ \sim -11.5^\circ$~\cite{Olive:2016xmw}. In this work, we will take $\theta_P
= -19.3^\circ$~\cite{Liu:2006dq}.

With the above Lagrangians, we can derive  the  transition amplitudes for  $\Upsilon(nS)(p_1) \to [B^{(*)}(q_1) {\bar B}^{(*)}(q_3)] B^{(*)}(q_2) \to  h_b(1P)(p_2) \eta(p_3)$ shown in Fig.~\ref{fig:loops}
\begin{eqnarray}
\mathcal{M}_{BB[B^*]}&=& \int\frac{d^4q_2}{(2\pi )^4} [-2g_{\Upsilon BB}\varepsilon_{1\mu} q_2^\mu][-g_{B^*BP}p_{3\nu}][g_{h_bB^*B}\varepsilon_{2\alpha}]\nonumber\\
&&\times\frac {i} {q_1^2-m_1^2} \frac {i(-g^{\nu\alpha}+{q^\nu_2q^\alpha_2}/{m^2_2)}}
 {q_2^2-m_2^2} \frac {i} {q_3^2-m_3^2} \mathcal F(m_2,q^2_2), \nonumber \\
\mathcal{M}_{BB*[B^*]}&=& \int\frac{d^4q_2}{(2\pi)^4} [g_{\Upsilon B^*B}\varepsilon_{\mu\nu\alpha\beta}p_1^\mu\varepsilon_1^\nu q_3^\alpha][-g_{B^*BP}p_{3\delta}][g_{h_bB^*B^*}\varepsilon_{\theta\phi\kappa\lambda}p_2^{\theta}\epsilon_2^\phi]\nonumber\\
&&\times\frac {i} {q_1^2-m_1^2} \frac {i(-g^{\delta\kappa}+{q^\delta_2q^\kappa_2}/{m^2_2})}
 {q_2^2-m_2^2} \frac {i(-g^{\beta\lambda}+{q^\beta_3q^\lambda_3}/{m^2_3})} {q_3^2-m_3^2} \mathcal F(m_2,q^2_2), \nonumber \\
\mathcal{M}_{B^*B[B^*]}&=& \int\frac{d^4q_2}{(2\pi)^4} [-g_{\Upsilon B^*B}\varepsilon_{\mu\nu\alpha\beta}p_1^\mu\varepsilon_1^\nu q_1^\alpha][g_{B^*B^*P}\varepsilon_{\theta\phi\kappa\lambda}p_3^{\phi}2q_2^\kappa][g_{h_bB^*B}\varepsilon_{2\delta}]\nonumber\\
&&\times\frac {i(-g^{\beta\theta}+{q^\beta_1q^\theta_1}/{m^2_1})} {q_1^2-m_1^2} \frac {i(-g^{\lambda\delta}+{q^\lambda_2q^\delta_2}/{m^2_2})}
 {q_2^2-m_2^2} \frac {i} {q_3^2-m_3^2} \mathcal F(m_2,q^2_2), \nonumber \\
\mathcal{M}_{B^*B^*[B]}&=& \int\frac{d^4q_2}{(2\pi)^4} [-g_{\Upsilon B^*B^*}\varepsilon_1^\mu (2q_{1\alpha}g_{\mu\nu}-q_{1\mu}g_{\alpha\nu}+q_{3\nu}g_{\mu\alpha}) ][g_{B^*BP}p_{3\beta}][g_{h_bB^*B}\varepsilon_{2\delta}]\nonumber\\
&&\times\frac {i(-g^{\nu\beta}+{q^\nu_1q^\beta_1}/{m_1^2})} {q_1^2-m_1^2} \frac {i}
 {q_2^2-m_2^2} \frac {i(-g^{\alpha\delta}+{q^\alpha_3q^\delta_3}/{m^2_3})} {q_3^2-m_3^2} \mathcal F(m_2,q^2_2), \nonumber \\
\mathcal{M}_{B^*B^*[B^*]}&=& \int\frac{d^4q_2}{(2\pi)^4} [-g_{\Upsilon B^*B^*}\varepsilon_1^\mu (2q_{1\alpha}g_{\mu\nu}-q_{1\mu}g_{\alpha\nu}+q_{3\nu}g_{\mu\alpha}) ][g_{B^*B^*P}\varepsilon_{\theta\phi\kappa\lambda}p_3^{\phi}q_2^\kappa] [g_{h_bB^*B^*}\varepsilon_{\beta\rho\sigma\delta} p_2^\beta\varepsilon_2^\rho] \nonumber\\
&&\times\frac {i(-g^{\nu\theta}+{q^\nu_1q^\theta_1}/{m_1^2})} {q_1^2-m_1^2} \frac {i(-g^{\alpha\delta}+{q^\alpha_2q^\delta_2}/{m^2_2})}
 {q_2^2-m_2^2} \frac {i(-g^{\lambda\rho}+{q^\lambda_3q^\rho_3}/{m^2_3})} {q_3^2-m_3^2} \mathcal F(m_2,q^2_2),
\end{eqnarray}
where $p_1$, $p_2$ and $p_3$ are the four momenta of the
initial state $\Upsilon(nS)$, final state $h_b(1P)$ and $\eta$, respectively. $\varepsilon_1$ and $\varepsilon_2$ are the polarization vector of $\Upsilon(nS)$ and  $h_b(1P)$, respectively. $q_1$, $q_3$ and $q_2$ are the four momenta of the
bottomed meson connecting $\Upsilon(nS)$ and $\eta$, the bottomed meson connecting $\Upsilon(nS)$ and $h_b(1P)$, and the exchanged bottomed meson, respectively.

In the triangle diagrams of Fig.~\ref{fig:loops}, the exchanged bottomed mesons are off shell. To compensate the offshell
effects and   regularize the ultraviolet  divergence~\cite{Locher:1993cc,Li:1996cj,Li:1996yn}, we introduce the monopole form factor,
\begin{equation}
\mathcal{F}\left( m_2,q_2^{2}\right) =\frac{\Lambda^2-m_2^2}{
\Lambda^2-q_2^2} \, ,
\end{equation}
where $q_2$ and $m_2$ are the momentum and mass of the exchanged bottomed meson, respectively.  The parameter $\Lambda \equiv m_2+\alpha \Lambda_{QCD}$ and the QCD energy scale $\Lambda_{QCD} = 220 \mathrm{MeV}$. The  dimensionless parameter $\alpha$, which is usually of order $1$,  depends on the  specific process.

%%%%%%%%%%%%%%%%%%%%%%%
\section{Numerical Results}
\label{sec:results}
%%%%%%%%%%%%%%%%%%%%%%%
%%%%%%%%%%%%%%%%%%%%%%%%%%%%%%%%%%%%%%%
\begin{table}[htb]
\begin{center}
\caption{The coupling constants of $\Upsilon(5S)$ interacting with $B^{(*)}{\bar B}^{(*)}$. Here, we list the corresponding branching ratios of $\Upsilon(5S)\to B^{(*)}{\bar B}^{(*)}$. }\label{tab:coupling cosntants}
 \begin{tabular}{ccccccccc}
 \hline
 Final state  & $\mathcal{B}(\%)$ & Coupling & Final state & $\mathcal{B}(\%)$ & Coupling & Final state & $\mathcal{B}(\%)$ & Coupling\\ \hline
 $B {\bar B}$ & $5.5$  & $1.76$  & $B {\bar B}^*+c.c.$ & $13.7$  & $0.14$ GeV$^{-1}$  & $B^* {\bar B}^*$ & $38.1$  & $2.22$ \\ \hline
 $B_s {\bar B}_s$ & $0.5$  & $0.96$  & $B_s {\bar B}_s^*+c.c.$ & $1.35$  & $0.10$ GeV$^{-1}$  & $B_s^* {\bar B}_s^*$ & $17.6$  & $5.07$ \\ \hline
\end{tabular}
\end{center}
\end{table}
%%%%%%%%%%%%%%%%%%%%%%%%%%%%%%%%%%%%%%%%%%%%%%%%%%%%%%%%%

With the experimental data on the decay width  of $\Upsilon(4S) \to B {\bar B}$~\cite{Olive:2016xmw},  the coupling constant $g_{\Upsilon(4S) BB}$ is  determined as  $g_{\Upsilon(4S) BB}=24.2$ which is comparable to the estimation in the vector meson dominance model. Since the mass of $\Upsilon(4S)$ is only above the $B {\bar B}$ threshold, the coupling constants $g_{\Upsilon(4S) B^*B}$ and $g_{\Upsilon(4S) B^*B^*}$ are determined as follows
\begin{eqnarray}
g_{\Upsilon(4S)B^*B} = \frac {g_{\Upsilon(4S)BB}}{\sqrt{m_{B^*} m_B}}\, , \quad g_{\Upsilon(4S)B^*B^*} = g_{\Upsilon(4S)B^*B} \sqrt {\frac{m_{B^*}} {m_B}} m_{{B^*}}\, .
\end{eqnarray}

For the coupling constants between $\Upsilon(5S)$ and $B^{(*)} {\bar B}^{(*)}$, we use the  experimental data on the decay width of  $\Upsilon(5S) \to B^{(*)} {\bar B}^{(*)}$~\cite{Olive:2016xmw}. The measured  branching ratios and the corresponding coupling constants are given  in Table~\ref{tab:coupling cosntants}.  One can see that the values determined
from the $\Upsilon(5S)$ data in Table~\ref{tab:coupling cosntants} are very small. This is  partly due to the fact that as a high-excited $b{\bar b}$ state, the wave function of $\Upsilon(5S)$ has a complicated node structure, and the coupling constants will be small if the
$p$ values of $B^{(*)} {\bar B}^{(*)}$ channels ($1060-1270$ MeV) are to those corresponding to the zeros in the amplitude~\cite{Meng:2008bq}. Since there is no experimental information
on $\Upsilon(6S)\to B^{(*)}{\bar B}^{(*)}$~\cite{Olive:2016xmw}, we choose the same values as  the $\Upsilon(5S)$ ones.

The coupling constants between $h_b(1P)$ and $B^{(*)} {\bar B}^*$  in
Eq.~(\ref{eq:h2}) are determined as
\begin{eqnarray}
g_{h_b BB^*} &=& -2g_1 \sqrt{m_{h_b} m_B m_{B^*}} \, ,  \ \ g_{h_b
B^* B^*} =2 g_1 \frac{m_{B^*}}{\sqrt{m_{h_b}}} \,  ,
\end{eqnarray}
where $g_1=-\sqrt{{m_{\chi_{b0}}}/{3}}/{f_{\chi_{b0}}}$.
$m_{\chi_{b0}}$ and $f_{\chi_{b0}}$ are the mass and decay constant
of $\chi_{b0}(1P)$, respectively~\cite{Colangelo:2002mj}, i.e.
$f_{\chi_{b0}} = 175 \pm 55$ MeV~\cite{Veliev:2010gb}.

In the chiral and heavy quark limits, the couplings between bottomed meson pair and
light pseudoscalar mesons have the following
relationships~\cite{Casalbuoni:1996pg},
\begin{eqnarray}
g_{{\cal B}^\ast {\cal B}^\ast P} = \frac{g_{{\cal B}^\ast {\cal B} P}}{\sqrt{m_{\cal B} m_{{\cal B}^\ast}}} =
\frac{2}{ f_\pi } g ,
\end{eqnarray}
where $f_\pi = 132$ MeV is the pion decay constant, and $g = 0.59$~\cite{Isola:2003fh}.

%%%%%%%%%%%%%%%%%%%%%%%%%%%%%%
%%%%%%%%%%%%%%%%%%%%%%%%%%%%%%
\begin{figure}[hbt]
\begin{center}
\includegraphics[width=0.49\textwidth]{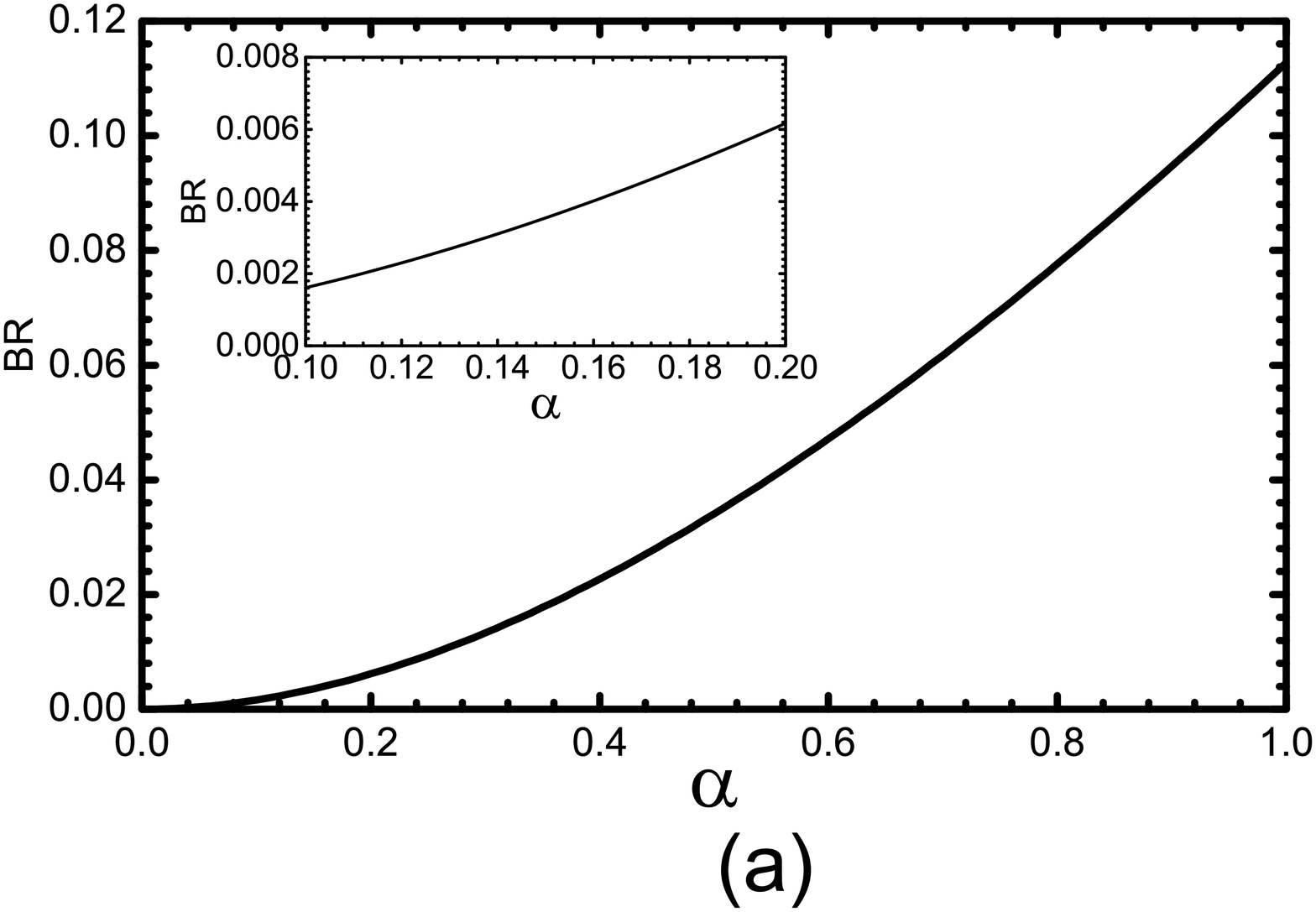}
\includegraphics[width=0.49\textwidth]{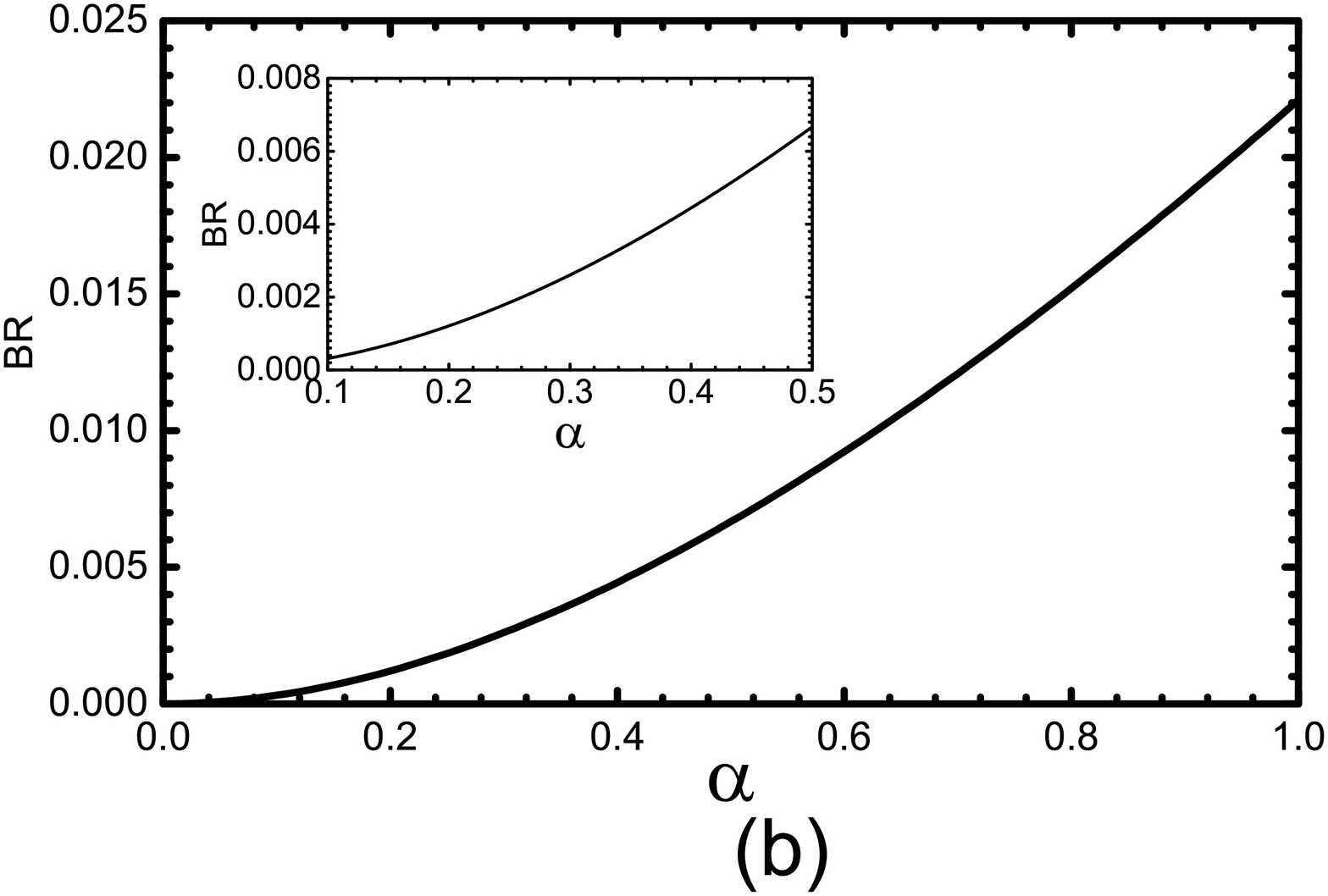}
\includegraphics[width=0.49\textwidth]{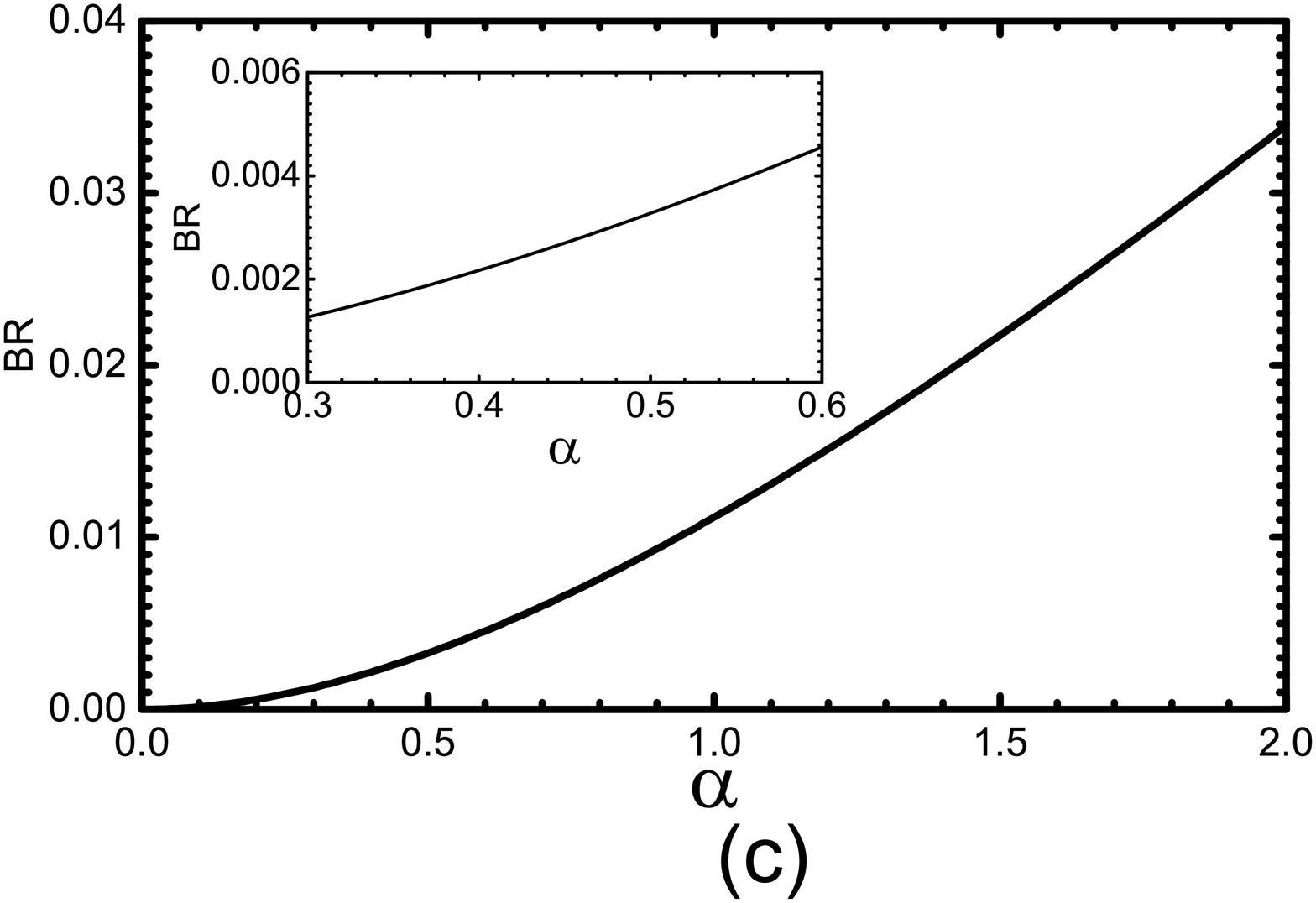}
\caption{(a). The $\alpha$ dependence of the branching ratios of $\Upsilon(4S) \to h_b(1P) \eta $. (b). The $\alpha$ dependence of the branching ratios of $\Upsilon(5S) \to h_b(1P) \eta $. (c). The $\alpha$ dependence of the branching ratios of $\Upsilon(6S) \to h_b(1P) \eta $.} \label{fig:BrOnAlpha}
\end{center}
\end{figure}
%%%%%%%%%%%%%%%%%%%%%%%%%%%%%%

%%%%%%%%%%%%%%%%%%%%%%%%%%%%%%
\begin{figure}[hbt]
\begin{center}
\includegraphics[width=0.49\textwidth]{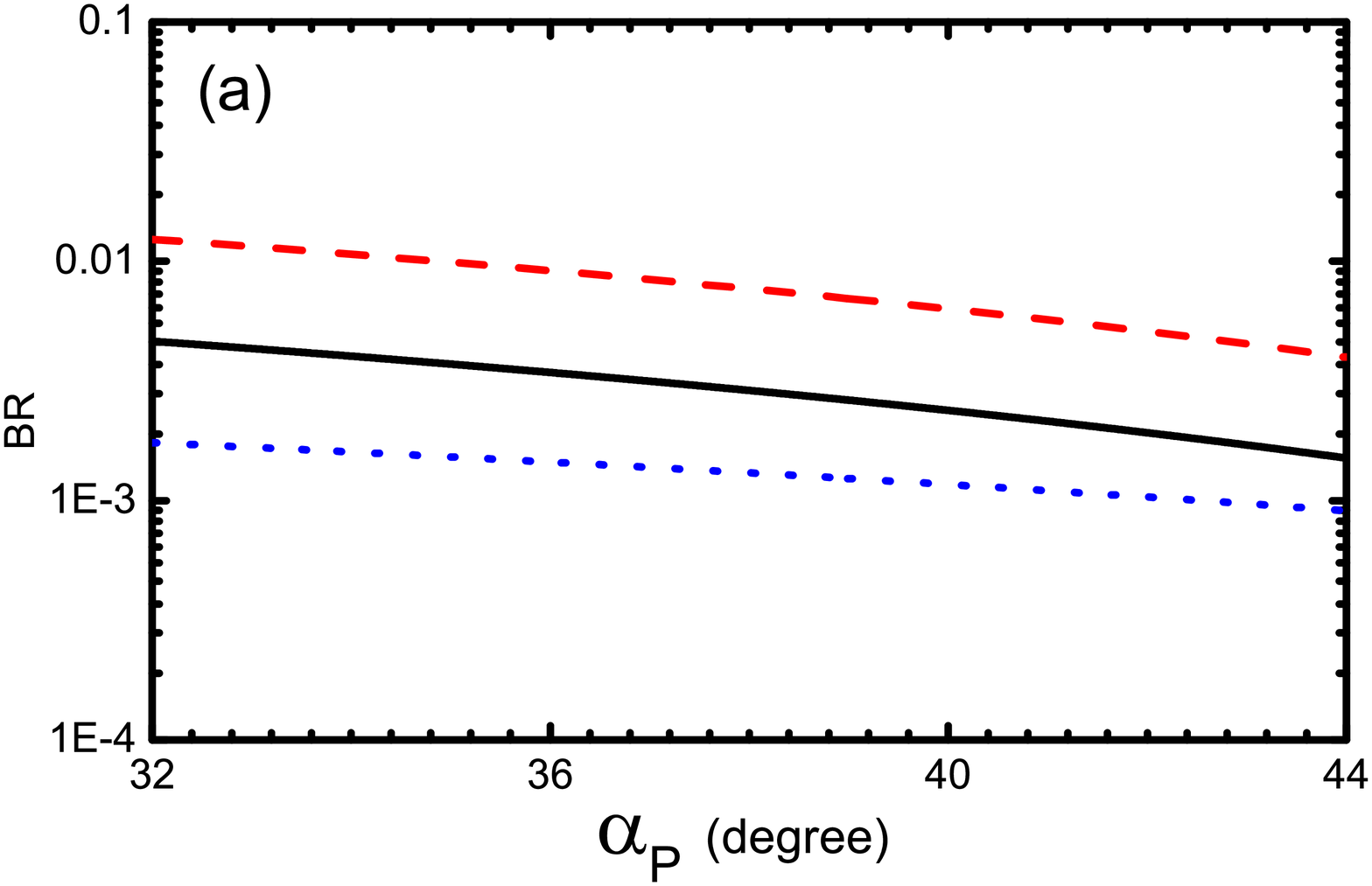}
\includegraphics[width=0.49\textwidth]{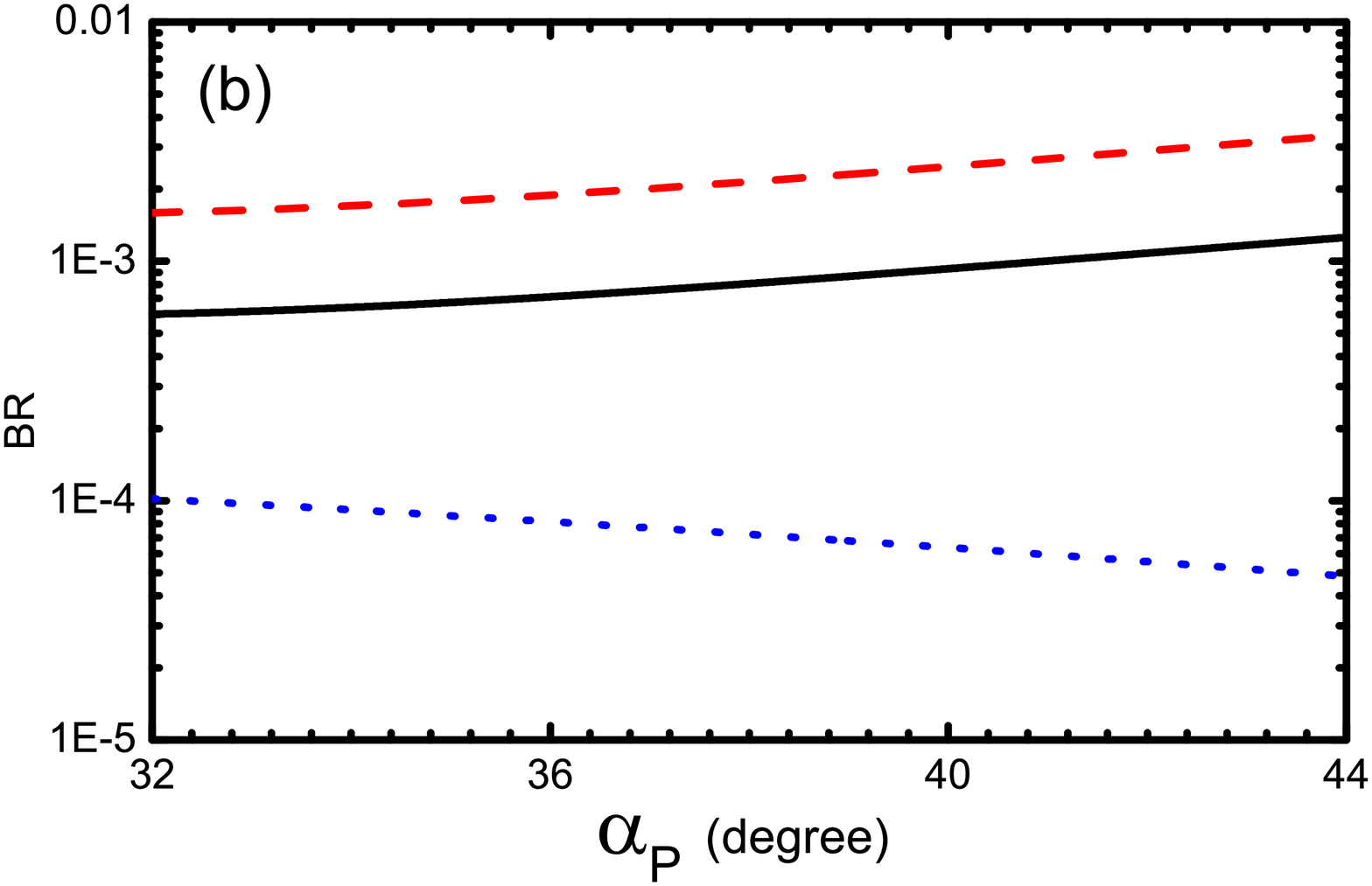}
\includegraphics[width=0.49\textwidth]{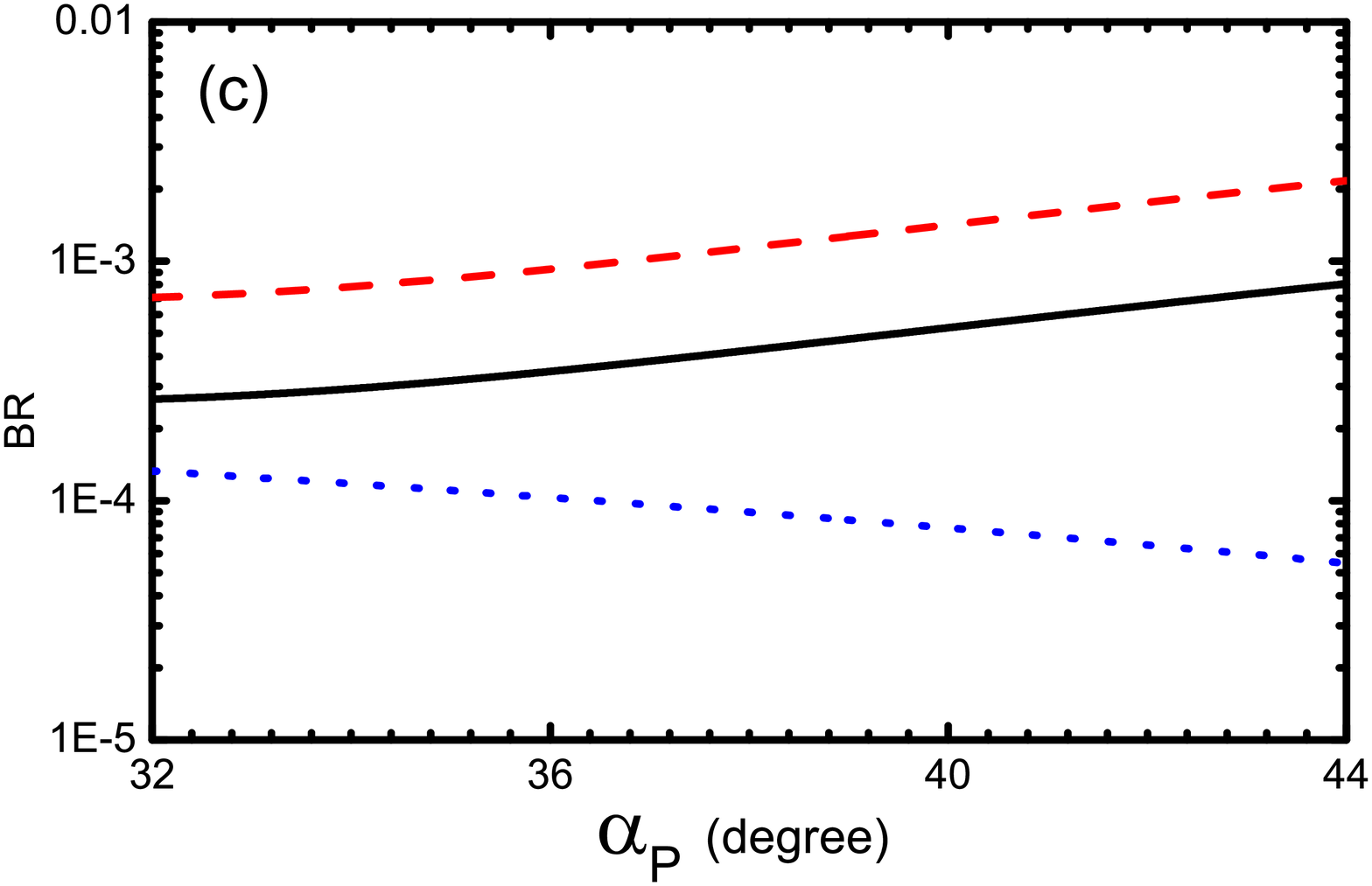}
\caption{(a).The dependence of branching ratios of $\Upsilon(4S) \to h_b(1P) \eta $ on the $\eta$-$\eta^\prime$ mixing angle with the cut-off parameter $\alpha=0.15$ (solid line)   and $0.25$ (dashed line), respcetively. The calculated branching ratios in NREFT approach are presented with dotted line. (b). The branching ratios of $\Upsilon(5S) \to h_b(1P) \eta $ in terms of the $\eta$-$\eta^\prime$ mixing angle with $\alpha=0.15$ (solid line) and $0.25$ (dashed line), respcetively. The calculated branching ratios in NREFT approach are presented with dotted line. (c). The branching ratios of $\Upsilon(6S) \to h_b(1P) \eta $ in terms of the $\eta$-$\eta^\prime$ mixing angle with $\alpha=0.15$ (solid line) and $0.25$ (dashed line), respcetively. The calculated branching ratios in NREFT approach are presented with dotted line.} \label{fig:BrOnmxing}
\end{center}
\end{figure}
%%%%%%%%%%%%%%%%%%%%%%%%%%%%%%
For the tree-level contributions to $\Upsilon(nS) \to h_b(1P) \eta$, the amplitude scales as the quark mass difference
\begin{eqnarray}
{\cal M}^{tree} \sim \delta
\end{eqnarray}
with $\delta=m_s-(m_u+m_d)/2$.

For the bottom meson loop contributions in Fig.~\ref{fig:loops}, the decay amplitude scales as follows,
\begin{eqnarray}\label{eq:power-counting-loop}
{\cal M}^{loop} \sim  {\cal N} \frac {q^2} {{\bar v}^3 M_B^2} \triangle \, ,
\end{eqnarray}
where ${\cal N}={1} /{(2{\sqrt 3}\pi v_b^4)}$, $q$ is the final $\eta$ momentum, ${\bar v}$ is understood as the average velocity of the
intermediate bottomed mesons. The meson mass difference $\triangle$ denotes the violation of the $SU(3)$ symmetry, which has similar size as $\delta$. $v_b$ denotes the bottom quark velocity inside the bottomonia and we take $v_b={\sqrt {0.1}}$ here.

For $\Upsilon(4S)\to h_b(1P)\eta$ decay, the momentum of the emitted $\eta$ is $q \simeq 388$ MeV and the velocity $v$ is about $\sqrt{[2m_B-(m_{\Upsilon(4S)}+m_{h_b})/2]/m_B}\simeq 0.28$. As a result, the factor ${\cal N}{ q^2} /({{\bar v}^3 M_B^2})$ is about $2.17$, which gives an enhancement compared with the tree-level contributions. For $\Upsilon(5S)\to h_b(1P)\eta$, the velocity ${\bar v} \simeq 0.23$ and $q=750$ MeV, so the factor ${\cal N}{ q^2} /({{\bar v}^3 M_B^2})$ is about $15$. For $\Upsilon(6S)\to h_b(1P)\eta$,  the velocity ${\bar v} \simeq 0.19$ and $q=930$ MeV, so the factor ${\cal N}{ q^2} /({{\bar v}^3 M_B^2})$ is about $37$. According our power counting analysis, the transitions $\Upsilon(4S,5S,6S) \to h_b(1P) \eta$ are dominated by the meson loops.

In Fig.~\ref{fig:BrOnAlpha} (a), we plot the branching ratios for $\Upsilon(4S) \to h_b(1P) \eta$ in terms of the cutoff parameter $\alpha$ with the monopole form factor. We also zoom into details of the figure with a narrow range $\alpha=0.1-0.2$ in order to show the best fit of the $\alpha$ parameter. As shown in Fig.~\ref{fig:BrOnAlpha} (a), the branching ratios are not drastically sensitive to the cutoff parameter $\alpha$. Our calculated branching ratios can reproduce the experimental data~\cite{Olive:2016xmw} at about $\alpha=0.12$. In Fig.~\ref{fig:BrOnAlpha} (b) and (c), we plot the predicted branching ratios for $\Upsilon(5S) \to h_b(1P) \eta$ and $\Upsilon(6S)\to h_b(1P)\eta$ in terms of the cutoff parameter $\alpha$ with the monopole form factor. The behavior is similar to that of $\Upsilon(4S) \to h_b(1P) \eta$ in Fig.~\ref{fig:BrOnAlpha} (a). The predicted branching ratios of $\Upsilon(5S) \to h_b(1P) \eta$ are about $10^{-3}\sim 10^{-2}$ with commonly accepted $\alpha$ range. For $\Upsilon(6S) \to h_b(1P) \eta$, the results are much small, which are about $10^{-4} \sim 10^{-2}$. At the same $\alpha$, the predicted branching ratios of $\Upsilon(5S) \to h_b(1P) \eta$ are about $1$ order of
magnitude smaller than that of $\Upsilon(4S) \to h_b(1P) \eta$. For $\Upsilon(6S) \to h_b(1P) \eta$, the predicted branching ratios are about $2$ orders smaller than that of $\Upsilon(4S) \to h_b(1P) \eta$. We suggest future experiment BelleII to carry out the search for the spin-flipping transitions $\Upsilon(5S, 6S) \to h_b(1P) \eta$ which will help us understanding the decay mechanism. Here, we should notice several uncertainties may influence our numerical results, such as the coupling constants
and off-shell effects arising from the exchanged particles of
the loops, and the cutoff parameter can also be different in
decay channels.

In order to illustrate the impact of the $\eta$-$\eta^\prime$ mixing angle, in Fig.~\ref{fig:BrOnmxing}, we present the branching ratios in terms of the $\eta$-$\eta^\prime$ mixing angle with $\alpha=0.15$ (solid line) and $0.25$ (dashed line), respectively. As shown in this figure, when the $\eta$-$\eta^\prime$ mixing angle $\alpha_P$ increases, the branching ratios of $\Upsilon(4S)\to h_b(1P)\eta$ decrease, while the branching ratios of $\Upsilon(5S,6S)\to h_b(1P)\eta$ increase.
This behaviour suggests how the $\eta$-$\eta^\prime$ mixing angle influences our calculated results to some extent.

 As a comparison, in Fig.~\ref{fig:BrOnmxing}, we also give the results using  the NREFT approach denoted as  dotted lines. The NREFT approach provides a systematic tools to control the uncertainties~\cite{Guo:2010ca,Guo:2010ak,Guo:2009wr}.  From the figure, one can see that for the decays $\Upsilon(4S) \to h_b(1P) \eta$, the NREFT results are at the same order of magnitude but smaller than the ELA results by  a factor of 2 to 5.  These differences may give some sense of the theoretical uncertainties for the calculated rates and indicates the viability of our model to some
extent. However, for transitions where the mass difference between the initial and final state becomes large, the NREFT may be not applicable. From Fig.~\ref{fig:BrOnmxing}, one can see that   for the decays $\Upsilon(5S, 6S) \to h_b(1P) \eta$  the NREFT results are smaller than the ELA results by approximately one order of magnitude. We suggest future experiment BelleII to carry out the search for this anomalous $\Upsilon(5S, 6S)\to h_b(1P) \eta$ transitions which will help us testing this point.

%%%%%%%%%%%%%%%%%%
\section{Summary}
\label{sec:summary}
%%%%%%%%%%%%%%%%%%
Recent experiments on the $\Upsilon(4S) \to h_b(1P) \eta$ transition show anomalously large decay rates. This seems to contradict the naive expectation that hadronic transitions with spin flipping terms should be suppressed with respect those that do not have these terms. In this work, we have studied  the spin-flipping transitions of $\Upsilon(4S, 5S, 6S)\to h_b(1P) \eta$ via intermediate bottomed meson loops in an effective Lagrangian approach. Our results have shown that the intermediate bottomed meson loops can play an important role in these process, especially when the initial
states are close to the two particle thresholds. For $\Upsilon(4S) \to h_b(1P)\eta$, the experimental data can be reproduced in this approach with a
commonly accepted range of values for the form factor cutoff parameter $\alpha$. We also predict the branching ratios of $\Upsilon(5S)\to h_b(1P)\eta$, which is about orders of $10^{-3}\sim 10^{-2}$. For $\Upsilon(6S) \to h_b(1P) \eta$, the results are much small, which are about $10^{-4} \sim 10^{-2}$. As a cross-check, we also calculated the
branching ratios of the decays in the framework of NREFT. We suggest future experiment BelleII to carry out the search for the spin-flipping transitions $\Upsilon(5S, 6S) \to h_b(1P) \eta$ which will help us understanding the decay mechanism.

%%%%%%%%%%%%%%%%%%
\section*{Acknowledgements}
\label{sec:acknowledgements}
%%%%%%%%%%%%%%%%%%

We would like to acknowledge Wei Wang for
carefully reading the manuscript and useful suggestions. This work is supported in part by the National Natural Science Foundation of China (Grant No. 11675091, No. 11575100, and No. 11505104).

\end{document}